\documentclass[preprint]{emulateapj}
\usepackage{hyperref}
\usepackage[usenames,dvipsnames]{color}
\usepackage{amsmath}
\newcommand{\Mstar}{M_{\star}} 
\newcommand{\Rstar}{R_{\star}} 
\newcommand{\Mdot}{\dot{M}} 
\newcommand{\Msun}{M_{\odot}} 
\newcommand{\Lstar}{\mathcal{L}_{\star}} 
\newcommand{\G}{\mathcal{G}} 

\newcommand{\be}{\begin{equation}}
\newcommand{\ee}{\end{equation}}
\def\lta{\,\raise 0.3 ex\hbox{$ < $}\kern -0.75 em
 \lower 0.7 ex\hbox{$\sim$}\,}
\def\gta{\,\raise 0.3 ex\hbox{$ > $}\kern -0.75 em
 \lower 0.7 ex\hbox{$\sim$}\,} 
\newcommand{\clength}{\lambda}
\newcommand{\thick}{\ell}

\newcommand{\Mdisk}{M_{\textrm{disk}}} 
\newcommand{\Ham}{\mathcal{H}} 
\newcommand{\Zstar}{\mathcal{Z}_{\textrm{star}}} 
\newcommand{\Zdisk}{\mathcal{Z}_{\textrm{disk}}} 
\newcommand{\zstar}{z_{\textrm{star}}} 
\newcommand{\zdisk}{z_{\textrm{disk}}} 

\newcommand{\etastar}{\eta_{\textrm{star}}} 
\newcommand{\etadisk}{\eta_{\textrm{disk}}} 
\begin{document}

\title{Alignment of protostars and circumstellar disks during the embedded phase}

\shorttitle{Early Stellar Alignment} 
\shortauthors{Spalding et al.} 
\author{Christopher Spalding$^{1}$, Konstantin Batygin$^{1}$, and Fred C. Adams$^{2,3}$} 

\affil{$^1$Division of Geological and Planetary Sciences, California Institute 
of Technology, Pasadena, CA 91125} 
\affil{$^2$Department of Physics, University of Michigan, Ann Arbor, MI 48109} 
\affil{$^3$Department of Astronomy, University of Michigan, Ann Arbor, MI 48109} 
\email{cspaldin@caltech.edu, kbatygin@gps.caltech.edu, fca@umich.edu}
\begin{abstract}
Star formation proceeds via the collapse of a molecular cloud core over multiple dynamical timescales. Turbulence within cores results in a spatially non-uniform angular momentum of the cloud, causing a stochastic variation in orientation of the disk forming from the collapsing material. In the absence of star-disk angular momentum coupling, such disk-tilting would provide a natural mechanism for production of primordial spin-orbit misalignments in the resulting planetary systems. However, owing to high accretion rates in the embedded phase of star formation, the inner edge of the circumstellar disk extends down to the stellar surface, resulting in efficient gravitational and accretional angular momentum transfer between the star and the disk. Here, we demonstrate that the resulting gravitational coupling is sufficient to suppress any significant star-disk misalignment, with accretion playing a secondary role. The joint tilting of the star-disk system leads to a stochastic wandering of star-aligned bipolar outflows. Such wandering widens the effective opening angle of stellar outflows, allowing for more efficient clearing of the remainder of the protostar's gaseous envelope. Accordingly, the processes described in this work provide an additional mechanism responsible for sculpting the stellar Initial Mass Function (IMF).
\end{abstract}
\section{Introduction}
\label{sec:intro} 

In the simplest picture for star and planet formation, the angular
momentum vectors for stellar rotation, the circumstellar disk, and the
resulting planetary orbits all coincide. However, recent observations showing that planetary orbits are often misaligned
with stellar rotation axes \citep{fabwinn,Winn2010} have prompted several authors (e.g. \citealt{Bate2010}; \citealt{Batygin2012}) to suggest that disks themselves may become misaligned with their parent stars. Any such primordial star-disk misalignment occurring within the embedded phase, during which the star gains most of its mass, has consequences both for future planetary systems and for the impact
of protostellar outflows on their surrounding envelopes. In this Letter, we construct a model for protostar-disk systems that describes the
gravitationally-facilitated precession of the stellar rotation
axis about a tilting disk, including dissipative torques owing to accretion.


In spite of enormous progress in our understanding of star formation
(from \citealt{sal87} to \citealt{mckeeost}), the final mass of a star
still cannot be unambiguously determined from the initial conditions
of the original molecular cloud core. Protostellar outflows represent
one mechanism that can help separate a newly formed star from its
immediate environment \citep{sal87}, and this mechanism may provide
an explanation for the stellar initial mass function \citep{Adams1996}.
Although outflows have sufficient mechanical luminosity to reverse the
infall \citep{ladaout}, one criticism of this picture is that the
outflows start with relatively narrow angular extents. However, the opening angles widen with time and precessing outflows can produce outflow cones that are effectively wider than their intrinsic extent, thereby making it easier for outflows to limit
the mass falling onto the central star/disk system. Independent of
the efficacy of the outflows in limiting stellar masses, observations
show that protostellar jets precess \citep{Eisloffel1996,cesaroni} and that
circumstellar disks are not always aligned with the plane of binary
orbits \citep{staple,koresko}.

The angular momentum of a circumstellar disk must be obtained from the gradual accumulation of material from a molecular cloud core. Rotation rates of such cores are estimated through measurements of velocity gradients of a given molecular line across the map of the
core (e.g., \citealt{Goodman1993}). The inferred angular velocity vectors \textit{do not point in the same direction over the entire core}; instead they
vary in projected direction over a range of $\sim30$ degrees within
the region encompassing material that is destined to form a star.
Moreover, the coherence length $\clength$ for the velocity vectors
inferred from these emission maps is $\clength\sim0.01$\,pc
\citep{caselli}. As collapse of these core structures proceeds, the
infalling material will thus sample cloud layers of differing angular momentum orientation. As the layers fall inward
and join the growing star/disk system, the angular momentum vector of
the system must vary in direction (as well as magnitude). On a
related note, these cores are observed to be turbulent, especially in
the outer layers of low-mass cores \citep{myerfuller} and in more
massive cores \citep{jijina}. The collapse of a turbulent region also
produces varying directions for the angular momentum vectors of the
forming star/disk systems as the collapse proceeds. Numerical
simulations of this process \citep{Bate2010,fielding} show that the
angular momentum vectors of the disks change as different cloud layers
fall inward.

Assuming the star to be decoupled from the disk, star-disk misalignment is indeed an expected result of disk-tilting. However, young stars rotate rapidly, becoming oblate. This oblateness leads to a gravitationally-forced precession of the stellar spin axis with respect to the disk (\citealt{Batygin2013}) and provides a physical mechanism by which the star's spin axis may trail the disk as it tilts. Additional potential sources of stellar spin-axis evolution include accretion, stellar winds and magnetic fields. As we are considering the Class 0 phase, accretion is likely to dominate over other effects, as discussed below.
\begin{figure*}
\includegraphics[trim=0cm 2cm 0cm 2cm, clip=true,width=1\textwidth]{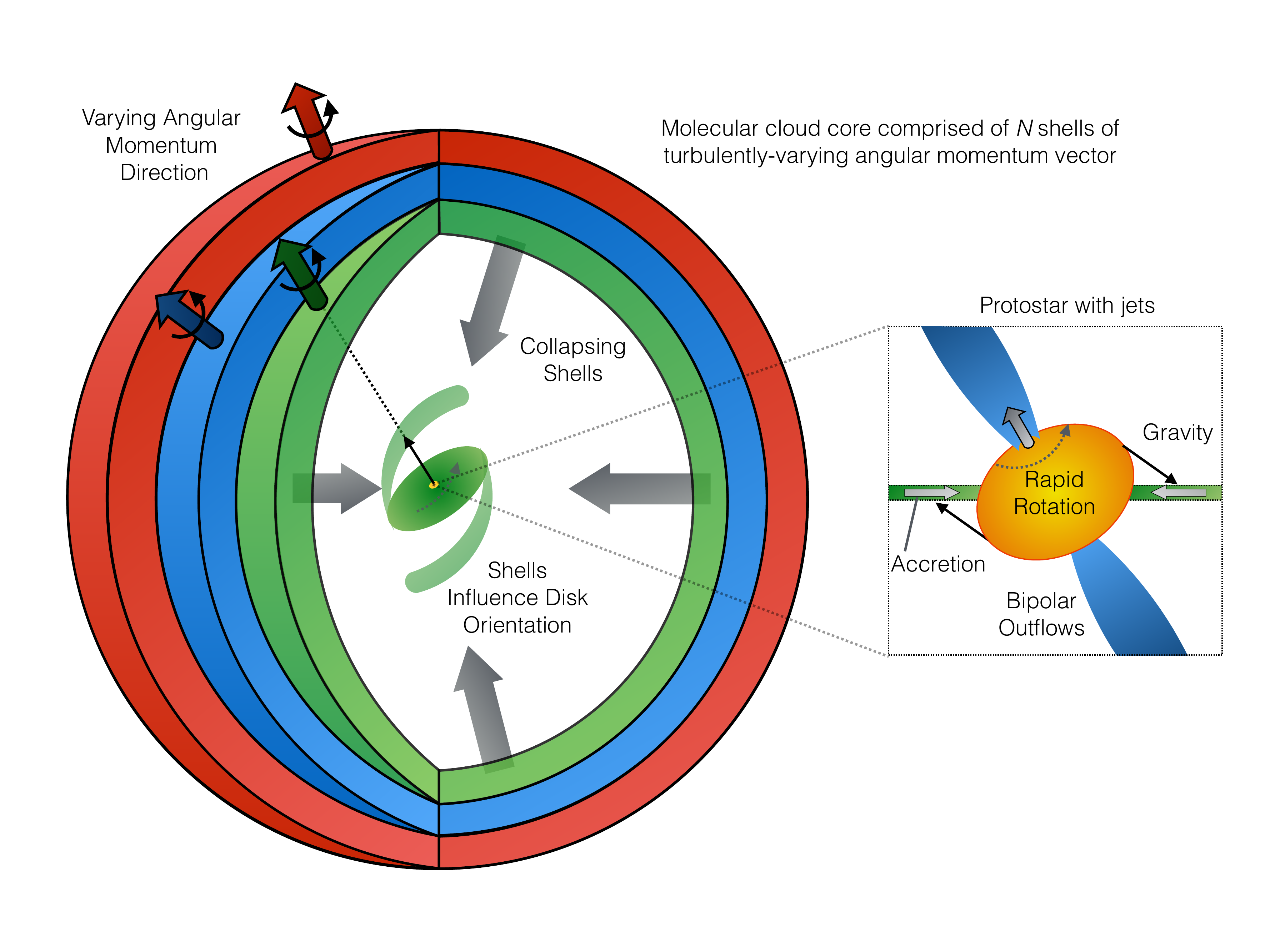}
\caption{A schematic of the process described in the text. Different shells possess different angular momentum vectors. In turn, the disk changes its orientation with time. Gravitational and accretional torques act between the star and disk, with bipolar outflows originating from the stellar spin axis.} 
\label{Schematic}
\end{figure*}

\section{Model Description} 
 We begin by describing the basic model, illustrated in Figure \ref{Schematic}, whereby a molecular cloud core collapses to form a star and circumstellar disk. As noted in the Introduction, sequential radii within the core differ in mean angular momentum direction. Owing to the large angular momentum reservoir of the collapsing material, the disk will in turn adopt a time-varying orientation as the core collapses onto it. Within the physical framework considered here, there exist three relevant timescales. Specifically, these are the shell-shell de-correlation time ($\tau_{\textrm{dc}}$), the characteristic star-disk realignment time ($\tau_{\textrm{r}}$) and the nodal regression period of the stellar spin-axis ($\mathcal{T}$). Let us evaluate these quantities sequentially. 
 \subsection{Turbulent Core Collapse}\label{1}
  Although the detailed structure of the core is complicated, we
assume that the density distribution has the form 
\be\rho(r)={A\,c_{\rm{s}}^{\,2}\over2\,\pi\,{G}\,r^2}\,,\label{rho}\ee
where $c_{\rm{s}}$ is the isothermal sound speed and $A>1$ is an overdensity
factor \citep{fatuzzo} that accommodates the fact
that cores are not in exact hydrostatic equilibrium \citep{lee}. With 
the density distribution (\ref{rho}), the enclosed mass has the form 
\be{M(r)}={2\,A\,c_{\rm{s}}^{\,2}\,r\over{G}}\,.\ee
Motivated by both observations of emission maps \citep{caselli} and
numerical simulations of collapsing turbulent cores
\citep{Bate2010,fielding}, we assume that different (spherical) shells
have different directions for their angular velocity. To be consistent
with observed maps and numerical expectations, the shell thickness
should be comparable to, but smaller than, the coherence length
$\clength$. For the sake of definiteness, we take the shell thickness
$\thick=\clength/2=0.005$ pc. With this choice, the formation of a
solar type star will involve the collapse of $N=5-10$ shells. The mass 
of each shell is given by 
\be\Delta{m}={2\,A\,c_{\rm{s}}^{\,2}\over{G}}\,\thick\approx0.10\,M_\odot 
\left({c_{\rm{s}}\over0.2\,\rm{km/s}}\right)^2 
\left({\thick\over0.005\,\rm{pc}}\right)\,.\ee
Each of these shells is then assumed to have an angular velocity
vector with direction chosen randomly within a range 0 -- 30 degrees.
 In this scenario, the mass infall rate is nearly constant with 
${\dot M} = m_0\, c_{\rm{s}}^3/G$, where $m_0$ is a dimensionless constant of
order unity \citep{shu77}. The shell-shell decorrelation time is equal to the corresponding time interval required
for a shell to fall inward, given by
\be \tau_{\textrm{dc}}={2A\over{m_0}}{\thick\over{c_{\rm{s}}}}\approx25,000\,\,\rm{yr}\,.
\label{mismatch}\ee
 In order to model angular velocity variation, we define scaled Poincar\'e action-angle coordinates in terms of inclination angle $\beta$ and longitude of ascending node $\Omega$:
 \begin{align}
\mathcal{Z} = 1-\cos(\beta) \ \ \ \ \ z =-\Omega.
\label{Poincare}
\end{align}
  We randomly choose the third Poincar\'e momentum of shell $i$ from a Gaussian distribution having mean $\mu=\mathcal{Z}_{i-1}$ and standard deviation $\sigma=\mathcal{Z}_{30^{\circ}}$\, where $\mathcal{Z}_{30^{\circ}}=1-\cos(30^{\circ})$ is the value of $\mathcal{Z}_{\textrm{disk}}$ corresponding to a 30\,degree inclination. Specifically,
 \begin{equation} \label{random}
 \mathcal{Z}_{i+1}=\mathcal{R}_{\rm{an}}\bigg[\exp{\bigg(-\frac{(\mathcal{Z}-\mathcal{Z}_{i})^2}{2 \mathcal{Z}_{30^{\circ}}^2}\bigg)}\bigg],
 \end{equation}
 \noindent where $\mathcal{R}_{\rm{an}}[]$ symbolises extracting a random number from the distribution within parentheses. Likewise, the canonical angle $z$ of shell $i$ relative to shell $i-1$ is drawn randomly, but from a uniform distribution of $0\leq-z\leq 2\pi$. Additional, small-scale turbulence-mediated stochasticity is introduced by way of $100\,N$ smaller shells, each varying in a Gaussian form by a value of 1\,degree. Once all shells have been prescribed a value of $\mathcal{Z}$ and $z$, we smooth the $100\,N$ inclinations into a continuous function of time by considering the entire star formation process to proceed over a time $\tau_{\textrm{form}}=2\times10^5$\,years and thus, for each of the 100\,$N$ inclinations to occur within a time interval of $\Delta t=\tau_{\textrm{form}}/100\,N$. Using a 3rd order cubic spline interpolation, we generate a pseudo-random function of time, \{$\mathcal{Z}(t),z(t)$\}, denoting the time-varying angular momentum vector of the collapsing material. Assuming the disk to obtain its angular momentum directly from the envelope, the disk angular momentum vector instantaneously follows that of the collapsing shells.
 \subsection{Realignment Time}
The angular momentum of a rotating star is given by the well-known expression
\begin{align}
\Lstar=I\,\Mstar\,\Rstar^2\,\omega,
\end{align}
where $I$ is the dimensionless moment of inertia and $\omega$ is the stellar spin. For the purposes of this Letter, we shall adopt stellar structure parameters corresponding to $n=3/2$ polytrope, which corresponds to a fully convective object.

At the earliest stages of stellar evolution, accretion rates of disk material onto the protostellar core can be ubiquitously high with characteristic values of order $\Mdot\sim 10^{-5} \, \Msun$/yr (for a $M\sim1\,\Msun$ object; see \citealt{WardThompson2002}). Accordingly, the accretionary ram pressure may in fact be sufficient to overwhelm the magnetic pressure of the protostellar magnetosphere, connecting the disk's inner edge to the stellar surface (\citealt{Gosh1978}). The critical magnetic field strength, $B_{\rm{crit}}$, below which this happens can be estimated by setting the magnetospheric disk truncation radius to that of the star (\citealt{Shu1994}):
\begin{align}
B_{\rm{crit}}=\bigg(\frac{G\,\Mstar\,\Mdot^2\,\xi^7}{\Rstar^5}\bigg)^{1/4},
\end{align}
where $\xi$ is a dimensionless constant of order unity (\citealt{Mohanty2008}). Given nominal parameters, the critical field evaluates to $B_{\rm{crit}} \sim 1$\,MG, which is well above the oft-cited $B_{\star} \sim 1$ kG fields inherent to young stellar objects (\citealt{Gregory2012}).

In a regime where the inner edge of the disk is physically connected to the stellar surface, the accretionary flow will facilitate a direct and efficient transfer of angular momentum between the disk and the host star (\citealt{Gosh1979}). While the details of the disk-star coupling in a shearing boundary layer can be complex (\citealt{Belyaev2013}), to leading order the rate of stellar angular momentum accumulation can be written as follows (\citealt{Armitage1996}):
\begin{align}
\frac{d\Lstar}{dt}\simeq\Mdot\,\sqrt{\G\,\Mstar\,\Rstar}.
\end{align}
\indent With the above equations at hand, we may now define a characteristic timescale for accretion-forced realignment of the stellar spin-axis. Specifically, we have:
\begin{align}
\tau_{\rm{r}}\equiv\frac{\Lstar}{\textrm{d}\Lstar/\textrm{d}t}\sim\frac{I\,\Mstar\,\Rstar^2\,\omega}{\Mdot\,\sqrt{\G\,\Mstar\,\Rstar}}\,\sim\,10^4\,\,\,\rm{years},
\end{align}
where as an estimate of the stellar spin rate we adopt the break-up rotational velocity $\omega = \sqrt{\G \Mstar / \Rstar^3}$, leading to $\tau_{\rm{r}}=I\,(\Mstar/\Mdot)$, which is independent of stellar radius, $\Rstar$. Note the similarity between $\tau_{\textrm{r}}$ and the shell infall timescale (equation~\ref{mismatch})

Additional effects can change the alignment. Perhaps most notably, modulation of the stellar spin-axis may arise from magnetic disk-star coupling (\citealt{Lai2011}; \citealt{Spalding2014}). While we have neglected this effect here, the fact that the accretionary flow at early stages of stellar evolution is intense enough to penetrate the stellar magnetosphere suggests that, indeed, the dominant mode of realignment will be facilitated by accretion and not magnetohydrodynamic effects.
\subsection{Precession}

As mentioned above, during the Class 0 epoch of stellar formation, young stellar objects may spin at near-breakup velocities. This naturally leads to significant rotational deformation. The spin-axis dynamics of an oblate spheroid can be modeled using standard techniques of celestial mechanics by replacing the rotational bulge of the star with an inertially equivalent orbiting ring of semi-major axis
\begin{align}
\chi=\left(\frac{16\,\omega^2\,k_2^2\,\Rstar^6}{9\,I^2\,\G\Mstar}\right)^3=\Rstar\left(\frac{4\,k_2}{3\,I}\right)^{2/3},
\end{align}
where $k_2=0.14$ is the Love number\footnote{This value corresponds to a polytropic body of index $n=3/2$} (twice the apsidal motion constant). The second equality follows from assuming that the star spins at breakup frequency. In principle, the aforementioned ring has a well-specified mass, however, its value only controls the back-reaction of the stellar quadrupole moment on the disk, which is unimportant.

To complete the specification of the problem, we must characterize the properties of the disk. We take the disk to be axisymmetric, and its surface density to vary inversely with semi-major axis (\citealt{Andrews2010}):
\begin{align}
\Sigma = \Sigma_0 \left(\frac{a}{a_0} \right)^{-1},
\end{align}
where $\Sigma_0$ is the surface density at $a = a_0$. Additionally, we take the disk aspect ratio to be $\zeta\equiv h/a=0.05$, though its actual value likely varies with disk radius up to $\sim0.1$ (\citealt{Armitage2011}). We note that under this prescription, 
\begin{align}
\Mdisk=\int_0^{2\pi}\int_{\Rstar}^{a_{\rm{out}}}\Sigma\,a\,da\,d\phi\simeq2\,\pi\Sigma_0\,a_0\,a_{\rm{out}},
\end{align}
where $a_{\rm{out}}=30\-- 50$ AU is the physical size of the disk (\citealt{Kretke2012,Anderson2013}).

To compute the dynamical evolution, we make use of classical perturbation theory (\citealt{Morbidelli2002}). Accordingly, we must first choose the appropriate expansion of the disturbing Hamiltonian. Given that $\chi \approx \Rstar$ and the inner boundary of the disk is linked to the stellar surface, an expansion in the semi-major axis ratio (\citealt{Kaula1962}; \citealt{Spalding2014}) is bound to be slowly-convergent. Therefore, in this work we shall opt for an alternative description that assumes mutual disk-star inclination as a small parameter and places no restrictions on the semi-major axis ratio (\citealt{LeVerrier1856},\citealt{Murray1999}). 

As a starting step, consider the mutual interaction of a massive hoop representing the stellar rotational bulge and a disk annulus of radial thickness $da$. It is a well-known result of secular perturbation theory that upon averaging over the orbital phase, the semi-major axes of both rings are rendered constants of motion. Thus, the Keplerian contributions to the Hamiltonian become trivial and can be omitted. 

To leading order in mutual inclination, the Lagrange-Laplace disturbing Hamiltonian reads (\citealt{Batygin2013}):
\begin{align}
d\Ham&=\frac{\tilde{b}_{3/2}^{(1)}}{4} \sqrt{\frac{\G\,\Mstar}{a^3}}\frac{dm}{\Mstar} \sqrt{\frac{\chi}{a}}\,\bigg[\Zstar \nonumber\\
 &-2\sqrt{\Zstar\Zdisk}\cos(\zstar-\zdisk)\bigg],\nonumber\\
\end{align}
where $dm=2\pi\,\Sigma_0\,a_0\,da$ is the mass of the perturbing annulus, $\tilde{b}_{3/2}^{(1)}$ is a softened Laplace coefficient (see below for an explicit expression).

 To obtain the Hamiltonian governing the interactions between the star and the full disk, we integrate with respect to the semi-major axis ratio $\alpha=\chi/a$:
\begin{align}
\Ham&=\frac{1}{4 \pi}\sqrt{\frac{\G\Mstar}{\chi^3}}\frac{\Mdisk}{\Mstar}\frac{\chi}{a_{\rm{out}}}\nonumber\\
&\times\left(\int_0^{\chi/\Rstar}\int_{0}^{2\pi}\frac{\cos(\psi)}{(1-2\alpha \cos(\psi)+\alpha^2+\zeta^2)^{3/2}}\,d\psi\,d\alpha\right)\nonumber\\
&\times\bigg[\Zstar-2\sqrt{\Zstar\Zdisk}\cos(\zstar-\zdisk)\bigg]\nonumber\\
\label{Hammy}
\end{align}
Note that in this formulation of the problem, we are not explicitly solving for the dynamical evolution of the disk using the above Hamiltonian. Instead, the time-varying variables $(\Zdisk,\zdisk)$ constitute prescribed functions of time, as described above. Suitably, the only equations of motion we derive from equation (\ref{Hammy}) are those corresponding to the $(\Zstar,\zstar)$ degree of freedom.

Although the inclination, $i$, and the longitude of ascending node, $\Omega$, are measured in an inertial reference frame, the inherent assumption of the Lagrange-Laplace secular theory is that the \textit{mutual} disk-star inclination remains small (\citealt{Morbidelli2012}). Thus, it is important to understand that any solution obtained within the framework of this description is only trustworthy if it dictates a low disk-star inclination for the entirety of the time-span of interest. Conversely, if mutual disk-star inclination is to increase to an appreciable value, one must default to the much more computationally expensive, but ultimately precise Gaussian averaging method (\citealt{Touma2009}).

To obtain the precession rate of the stellar spin axis in the frame of the disk, we may envision that the disk remains stationary at $\beta=0$ (this assumption will be lifted later), meaning that $\Zdisk=0$. This puts the amplitude of the harmonic part of the Hamiltonian (\ref{Hammy}) to zero, such that $\Ham$ governs pure rotation in $\zstar$. Accordingly, we have:
\begin{align}
\mathcal{T}=2\,\pi\left(\frac{\partial\Ham}{\partial\Zstar}\right)^{-1}\simeq130\left(\frac{0.01\Msun}{\Mdisk}\right)\left(\frac{\Mstar}{1\Msun} \right)\,\rm{years}.
\label{Tprec}
\end{align}
\indent For all reasonable choices of parameters, the precession timescale of the stellar spin-axis (which acts as the dynamical timescale of the problem at hand) is substantially shorter than both the accretionary realignment timescale and the shell-shell decoherence timescale. This feature is of crucial importance to understanding the results that follow, as it effectively guarantees that the dynamical evolution occurs within the \textit{adiabatic} regime, within which the star trails the disk's orientation.
\section{Numerical Simulations}
Equations of motion arising from Hamiltonian (\ref{Hammy}), as formulated in terms of action-angle coordinates (\ref{Poincare}) contain a coordinate singularity at $\Zstar=0$. This complication can be removed with a canonical change of variables. Specifically, we introduce a complex coordinate
\begin{align}
\eta=\sqrt{\mathcal{Z}}\cos(z)+\imath\sqrt{\mathcal{Z}}\sin(z),
\end{align}
where $\imath=\sqrt{-1}$. The Hamiltonian now reads:
\begin{align}
\Ham&=\mathcal{S}\left(\etastar\etastar^*+\etastar\etadisk^*+\etastar^*\etadisk\right),
\label{Hammy2}
\end{align}
where $\mathcal{S}=2\pi/\mathcal{T}$ is the coefficient on the first line of equation (\ref{Hammy2}) and the asterisk denotes a complex conjugate.

In addition to the dynamical evolution governed by $\Ham$, it is important to account for the dissipative effects originating from the realigning influence of accretionary torques, in the equations of motion. For tractability, it is sensible to parameterize such realignment as an exponential decay of the action\footnote{Introduction of such terms into the equations of motion tends to transform nearby elliptical fixed points into attractors (see e.g. \citealt{Batygin2011}.} $\Zstar$. Cumulatively, the relevant equation of motion takes the form:
\begin{align}
\frac{d\etastar}{dt}&=\imath\left(\frac{\partial\Ham}{\partial \etastar^*}\right)+\left(\frac{d\etastar}{dt}\right)_{\rm{r}}\nonumber\\
&=\imath\,\mathcal{S}\left(\etastar+\etadisk\right)-\frac{\etastar}{2\tau_{\textrm{r}}},\nonumber\\
\label{eta}
\end{align}
\noindent where ($d\etastar/dt )_{\rm{r}}$ describes the dissipative term, which acts to damp any misalignment. Without the dissipative term, equation~\ref{eta} describes conservative, gravitational precession of the stellar spin axis about a time-varying disk angular momentum vector.

To complete the specification of the problem, we prescribe the time evolution of a 1\,M$_{\odot}$ star as
\begin{equation}
M_{\rm{star}}(t)=\textrm{M}_{\odot}\bigg(\epsilon+\frac{t}{\tau_{\textrm{form}}}\bigg)
\end{equation}
 where $\epsilon=0.01$ represents a small initial `seed' mass onto which shells collapse. Additionally, the circumstellar disk mass ($M_{\rm{disk}}$) grows proportionally to that of the star such that 
 \begin{equation}
M_{\rm{disk}}(t)=0.1\,M_{\rm{star}}(t),
\end{equation}
\noindent with the 0.1 prefactor corresponding approximately to the upper limit for dynamical stability (\citealt{Armitage2011}). We consider a constant stellar radius of $\Rstar=4\,\rm{R}_{\odot}$ throughout (\citealt{Stahler1980}).  
\section{Results \& Discussion}

\begin{figure}[h!]
\centering
\includegraphics[trim=4cm 2cm 4.8cm 0cm, clip=true,width=1\columnwidth]{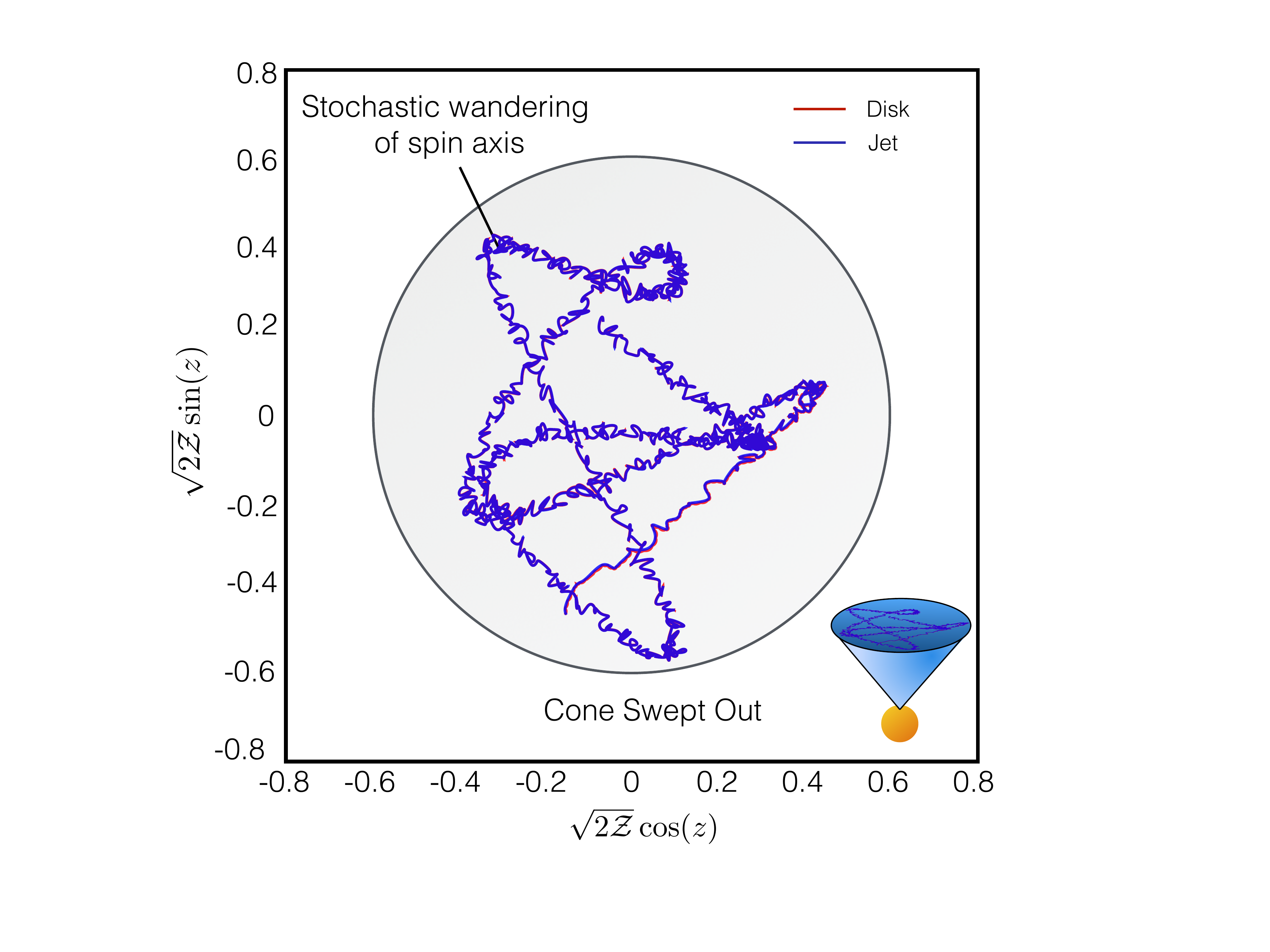}
\caption{The paths traced out by the angular momentum vectors of the disk (red) and star (blue) plotted in canonical Cartesian co-ordinates (see text). Notice that the red and blue paths almost exactly overlap. The shaded region approximately inscribes the cone of gas cleared out by stellar spin axis-aligned jets.} 
\label{SpinPath}
\end{figure}

\begin{figure}[h!]
\centering
\includegraphics[trim=0cm 2.6cm 0cm 4cm, clip=true,width=1\columnwidth]{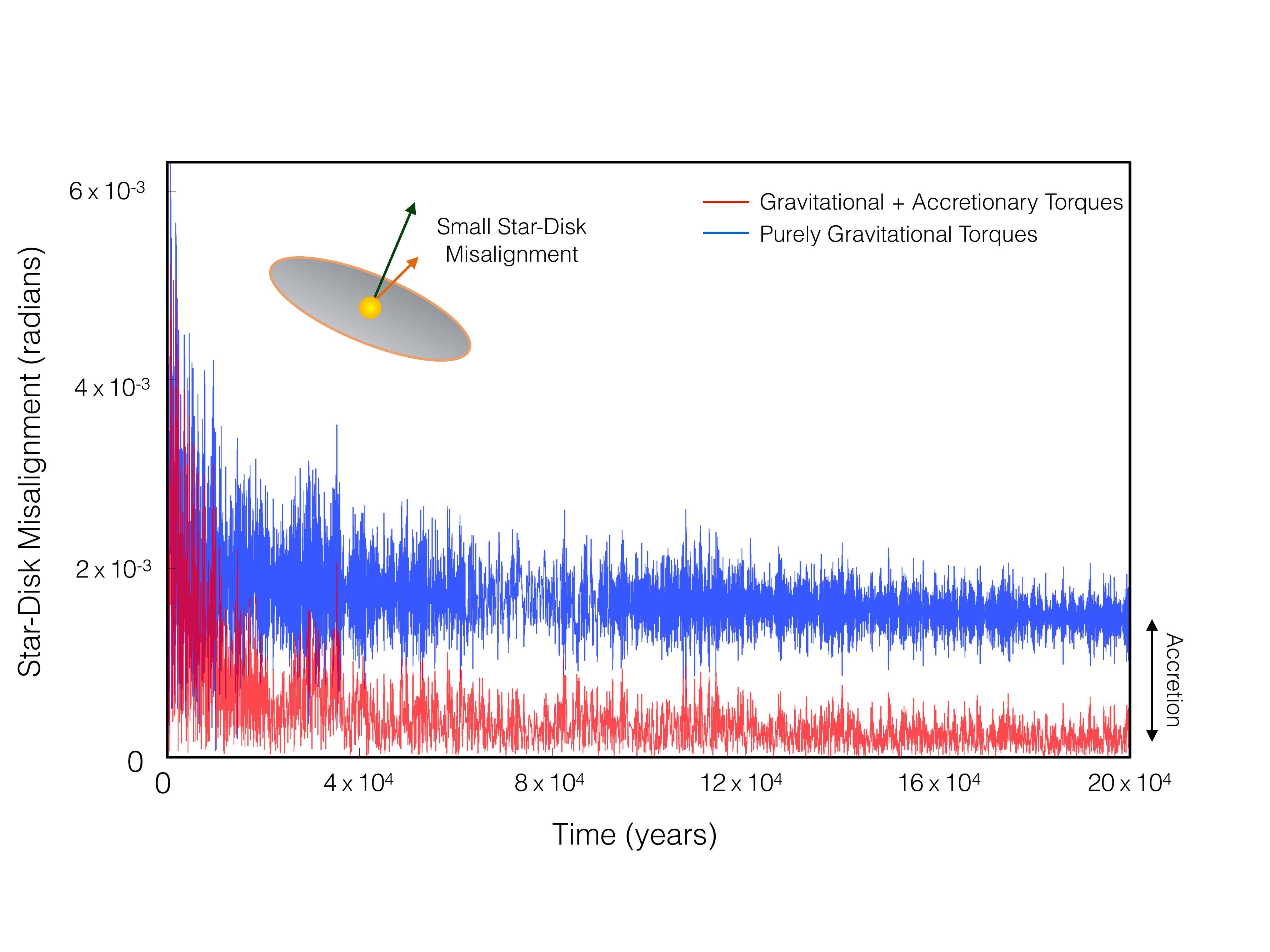}
\caption{The misalignment between star and disk angular momenta plotted as a function of time. Gravitational interactions alone (blue) are sufficient to suppress significant misalignment. Accretionary torques (red) further reduce the misalignment to near-zero values.}
\label{Aligned}
\end{figure}

In Figure \ref{SpinPath} we present the paths followed by the stellar and disk angular momentum vectors in the purely gravitational regime, i.e., zero accretion. As is immediately obvious, the two paths are indistinguishable, meaning that even in the absence of accretionary realignment, no significant star-disk misalignment can result from turbulent core collapse. Accretionary torques simply act to reduce the already-miniscule misalignments (Figure \ref{Aligned}) and so are dynamically unimportant to the problem at hand. As such, the first crucial result is that \textit{the hypothesis that turbulent core-collapse leads to primordial spin-orbit misalignments is inconsistent with the framework presented here}.  Spin-orbit misalignments must be obtained at a later evolutionary stage, such as during the main phase of planet formation (\citealt{Lai2011}; \citealt{Batygin2012}; \citealt{Batygin2013}; \citealt{Spalding2014}) or after the disk has dispersed (e.g., \citealt{Wu2011}; \citealt{Beauge2012}; \citealt{Albrecht2012}). 

Class 0 and Class I protostars possess collimated bipolar jets with sufficient mechanical luminosity to reverse the infall of core material. Such jets have been observed to `wiggle' in such a way as to suggest time evolution of the jet direction (\citealt{Eisloffel1996,cesaroni}). Previous pictures considering only disk motion do not necessarily account for jet wiggles as the jets are collimated along the stellar spin axis through the action of magnetic fields (\citealt{Shu1994}). Thus the jet itself is unlikely to move significantly if the star itself is not changing orientation. As noted by \citet{sal87}, a star breaks free of its enshrouding molecular envelope once outward pressures owing to stellar winds and jets exceed the ram pressure of infalling gas. Stellar outflows contribute significantly to such outward pressures and thus may in part determine the final mass of the forming star. Here we find that the stellar outflows carry out a random walk (Figure \ref{SpinPath}), leading to an effectively wider opening angle of the outflow. Accordingly, the wandering outflows may help separate the newly formed star/disk system from its environment earlier than would a stationary outflow. Such a physical process adds an important correction onto previous theories of star formation (e.g., \citealt{Adams1996}) which propose that the IMF may be determined in part by outflows. 

Here, we considered a constant mass infall rate of $\Mdot\approx10^5\,\Msun\,\rm{year}^{-1}$ over the entire star formation process. Strictly speaking, this assumption contradicts the non-steady nature of turbulent collapse. In principle, sufficiently violent episodic mass infall may lead to more rapid variations in disk angular momentum than considered here. However, owing to the vast separation of timescales between shell-shell decorrelation ($\tau_{\rm{dc}}\sim10^4\,\rm{years}$) and stellar precession ($\mathcal{T}\sim10^2\,\textrm{years}$) the star shall trail the disk under almost any reasonable collapse conditions. Additionally, the mass infall and accretion rates fall by about an order of magnitude over a longer timescale, between the Class 0 and Class I phases of star formation (\citealt{WardThompson2002}). Such a drop in infall rate is not included in our model but during the Class I phase, most of the mass is in the star-disk system (by definition) and so, combined with a drop in infall rate, disk tilting is likely to become significantly lower in amplitude. Accordingly, the star should remain even more tightly coupled to the disk, despite reduced accretionary torques, which we determined to be dynamically unimportant. 

This Letter presents a simple model for star/disk formation in molecular cloud cores possessing non-uniform angular momentum directions. We find that outflows change direction substantially, but stars and disks remain nearly aligned. Future work should develop more detailed models for all aspects of this problem, including the initial conditions, disk formation, wandering of outflow directions and misalignment between star and disk.\\

\noindent \textbf{Acknowledgements} 

 We would like to thank Richard Nelson for
useful discussions and the Michigan Institute for Research in
Astrophysics for helping to facilitate this collaboration.

 \end{document}